\documentclass[runningheads]{llncs}
\usepackage{graphicx}
\usepackage[symbol]{footmisc}
\usepackage{amsmath}
\usepackage{amssymb}
\usepackage{booktabs}
\usepackage{float}

\begin{document}
\title{T1/T2 relaxation temporal modelling from accelerated acquisitions using a Latent Transformer}
\titlerunning{T1/T2 relaxation modelling with Latent Transformer module}
%
\newcommand*\samethanks[1][\value{footnote}]{\footnotemark[#1]}
\author{
Fanwen Wang\samethanks\inst{1,2}\orcidID{0000-0001-5491-8333} 
\and
Michael T\"{a}nzer\thanks{Equal contribution}\inst{1,2}\orcidID{0000-0002-9046-1008} \and
Mengyun Qiao\inst{1}\orcidID{0000-0002-5157-1079} \and
Wenjia Bai\inst{1}\orcidID{0000-0003-2943-7698} \and
Daniel Rueckert\inst{1,3}\orcidID{0000-0002-5683-5889} \and
Guang Yang\inst{1,2}\orcidID{0000-0001-7344-7733} \and
Sonia Nielles-Vallespin\inst{1,2}\orcidID{0000-0003-0412-2796}
}
%
\authorrunning{Wang \& T\"{a}nzer et al.}
%
\institute{
Imperial College London, Exhibition Road, London SW7 2AZ, UK \and
Royal Brompton and Harefield hospital, Sydney Street, London SW3 6NP, UK \and
Technische Universität München (TUM), Arcisstraße 21, 80333 München, Germany
}

%
\maketitle              
\begin{abstract}
Quantitative cardiac magnetic resonance T1 and T2 mapping enable myocardial tissue characterisation but the lengthy scan times restrict their widespread clinical application. We propose a deep learning method that incorporates a time dependency \textit{Latent Transformer} module to model relationships between parameterised time frames for improved reconstruction from undersampled data. The module, implemented as a multi-resolution sequence-to-sequence transformer, is integrated into an encoder-decoder architecture to leverage the inherent temporal correlations in relaxation processes. The presented results for accelerated T1 and T2 mapping show the model recovers maps with higher fidelity by explicit incorporation of time dynamics. This work demonstrates the importance of temporal modelling for artifact-free reconstruction in quantitative MRI.
\end{abstract}
\section{Introduction}
Magnetic resonance imaging (MRI) is a crucial non-invasive tool for assessing tissue properties and functions, with applications across medical disciplines. In cardiac imaging, quantitative T1 and T2 mapping provide insights into myocardial composition and structure, enabling characterisation of cardiomyocytes~\cite{messroghli2007myocardial,sado2013identification,moon2013myocardial}. However, long scan times limit the clinical utility of cardiac T1 and T2 mapping. Although undersampled acquisitions offer a means to accelerate scans, they often lead to artifacts and errors, not only in the reconstructed images but also to a greater extent in the computed T1/T2 maps.

Recent deep learning approaches have shown promise for reconstructing high-quality maps from highly accelerated scans. Encoder-decoder models like AUTOMAP~\cite{AUTOMAP} leveraged deep convolutional neural networks to learn efficient representations directly from undersampled k-space and mapping targets. Schlemper et al.~\cite{DCNN} explored the ability of cascaded CNN to learn the spatial-temporal correlations from multi-coil undersampled cardiac cine MRI. Qin et al.~\cite{qin2021complementary} exploited the spatiotemporal correlations using recurrent network for dynamic multi-coil cardiac cine data. Lyu et al.~\cite{lyu2023region} divided temporal cine MRI data into several views and used a video-Transformer~\cite{yan2022multiview} model to capture spatial and temporal relationship. However, most existing methods disregard dependencies between parameterised time frames. As relaxation processes induce temporal correlations, explicitly modelling time structure is essential for accurate reconstruction.

We introduce an innovative deep reconstruction model that introduces a temporal dependency module to effectively capture inter-frame relationships within encoded sequences. The module is seamlessly integrated into an encoder-decoder architecture by modifying the latent vectors in the skip connections of the encoder-decoder model to better exploit the temporal correlation. By incorporating temporal dynamics, the proposed model aims to significantly enhance reconstructions derived from accelerated cardiac T1 and T2 mapping scans, regardless of the number of time points or mapping methodologies employed. This could facilitate accurate analyses of myocardial tissue properties from faster, patient-friendly scans. We present preliminary validation of our technique for accelerated cardiac T1 and T2 mapping against state-of-the-art methods and gold-standard fully-sampled acquisitions.

\section{Materials and Methods}

\subsection{Data Acquisition}
We used both single and multi-coil T1 and T2 mapping data from the MICCAI 2023 CMRxRecon training dataset~\cite{wang2023cmrxrecon}. The T1 mapping data was acquired using a Modified Look-Locker Inversion recovery (MOLLI) sequence \cite{messroghli_modified_2004} with nine frames of variable T1 weighting in short-axis view at end-diastole. For each subject, between five and seven slices were collected with a slice thickness of 5.0 mm. The matrix size of each T1-weighted frame was $144 \times 512$ with an in-plane spatial resolution of 1.4 x 1.4 mm$^2$. For the multi-coil data, the coils were compressed to 10 virtual coils. The T2 mapping data was acquired using a T2-prepared (T2prep) FLASH sequence with three T2 weightings and with geometrical parameters identical to the T1 mapping acquisition.

\subsection{Data processing}
Both single and multi-coil T1 and T2 mapping data from the 120 healthy subjects in the training dataset were randomly split into 80\% for training, 10\% for validation and 10\% for testing. We pre-processed the provided k-space data by scaling it to a range where the model could perform optimally. Specifically, we multiplied all k-space data by a fixed value of $10^2$ to bring the magnitude of the images values approximately into the [0, 1] range. This transformation is reversed before computing the mappings.

During training, the data was also augmented using a random undersampling mask for every subject. The random mask was generated by selecting every $k^{\textrm{th}}$ line starting from line $s$, where $k$ is the acceleration factor and $s$ is a randomly sampled integer between 0 and $k$. As in the original acquisition, the random mask preserved the central 24 lines of k-space. This allowed the model to exhibit greater adaptability to minor variations in the acquisition protocol, thereby providing a valuable and realistic data augmentation strategy.

\subsection{Model}
\subsubsection{Latent transformer}
The proposed model, the Latent Transformer (LT), employs an encoder-decoder architecture with shared encoding-decoding blocks across all time-frames (Fig.~\ref{fig:model-diagram}). Embedded within each skip connection between an encoding layer and the corresponding decoding layer is an LT block, which enables modelling of dependencies across frames before passing signals to the decoder layers (Fig.~\ref{fig:model-diagram}, E). Specifically, there is a unique LT block for each layer of the main encoder-decoder network. Each LT block utilises multi-layer and multi-head self-attention (Fig.~\ref{fig:model-diagram}, C and D) to compute the updated latent code as a weighted linear combination of itself and the latent codes of the other time-frames in a pixel-wise manner (Fig.~\ref{fig:model-diagram}, A and B). The LT blocks are key to exploit temporal correlations within the data to aid reconstruction performance, as they allow information from each frame to affect the reconstruction of all other frames. 

\subsubsection{Single-coil}
\begin{figure}[tbh]
    \centering
    \includegraphics[width=\linewidth]{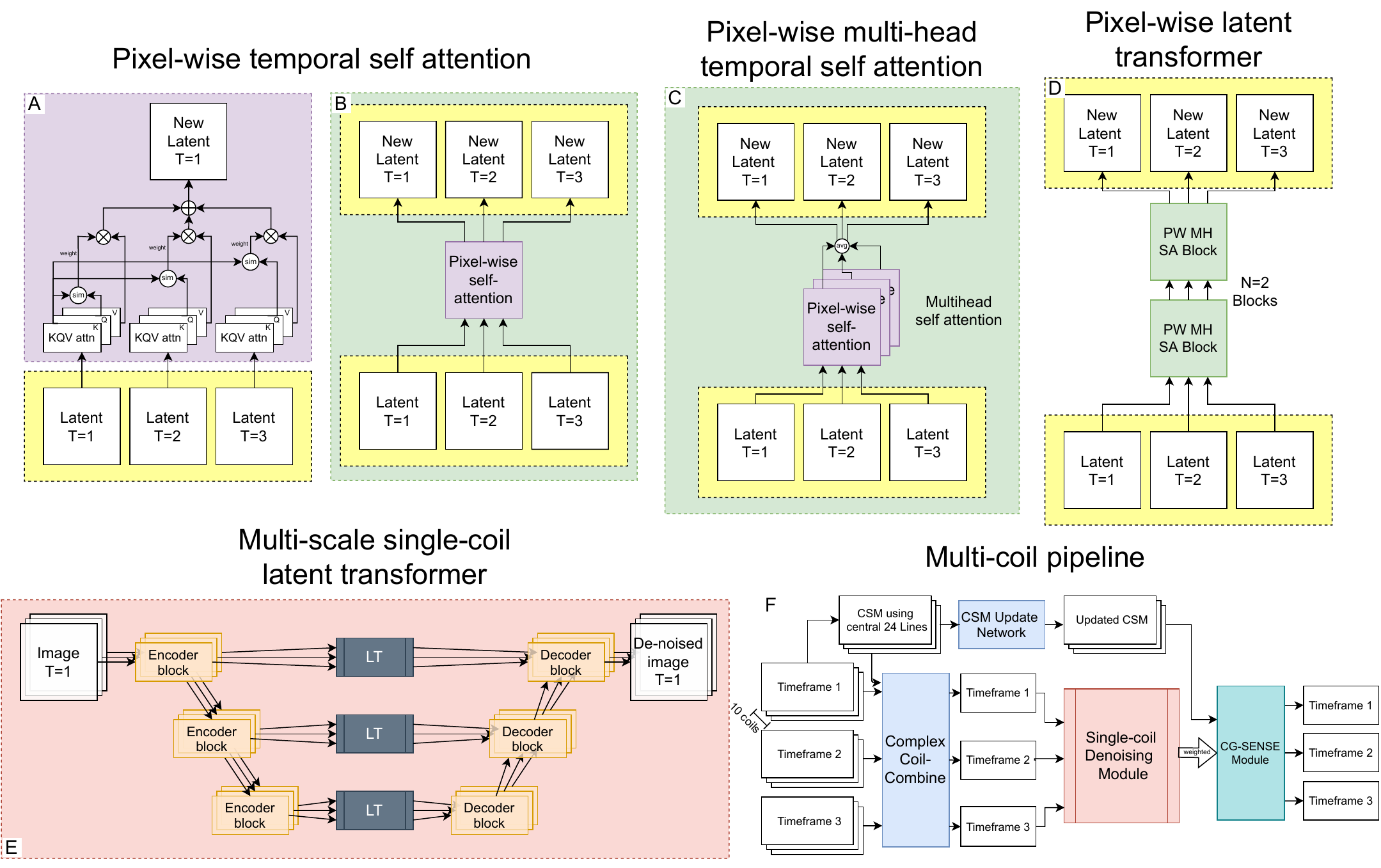}
    \caption{Diagram representing the architecture for the single-coil (E) and multi-coil (F) tasks. The figure also shows how the latent transformer is implemented using a pixel-wise self-attention mechanism (A, B) in a multilayer and multi-head fashion (D). The figure shows a case where three time-frames are available, but the method extends seamlessly to any number of time-frames with no modification.}
    \label{fig:model-diagram}
\end{figure}

The proposed model architecture consists of two complementary components as illustrated in Figure~\ref{fig:model-diagram}:
\begin{itemize}
\item An encoder-decoder network that serves as the main artifact removal model. In our experiments, we utilise a U-Net architecture as a strong baseline model. The goal of this encoder-decoder network is to remove large-scale artifacts from the input MRI frames.
\item The latent transformer model to explicitly capture inter-frame dependencies. 
\end{itemize}
Finally, the predictions from the main encoder-decoder model and the LT model are combined to produce the final reconstructed output frames. By fusing the outputs this way, the model leverages both general artifact removal capabilities and inter-frame dependencies for enhanced MRI reconstruction.

\subsubsection{Multi-coil model}
Similar to the single-coil model, the multi-coil artifact-removal architecture consists of a main encoder-decoder network and LT model tailored for multi-coil data. To reduce computational complexity, the LT is applied on the coil-combined complex image rather than each coil individually. Coil sensitivity maps (CSMs) $C$ are extracted using an iterative approach~\cite{CSMGenerate} with smoothing based on the central 24 lines of undersampled k-space $x \in {\mathbb{C}^{W \times M}}$ among all $W$ coils. The undersampled multi-coil data is multiplied by the conjugate CSMs to maintain complex information as $\hat{x} \in {\mathbb{C}^{W}}$ before input to the model.

The multi-coil reconstruction can be formulated as:
\begin{equation}
\hat{x}_{rec} = \mathop {\arg \min }\limits_{x} (\left\| {Ex - y} \right\|_2^2 + \lambda \left\| {\hat{x} - S(\hat{x};\theta)} \right\|_2^2)
\label{eqn:multi-coiloptimisation}
\end{equation}
\begin{equation}
E = M \cdot F \cdot \hat{C}
\end{equation}
where $x$ is the multi-coil complex image and $y\in {\mathbb{C}^{W \times M}}$ is the acquired multi-coil k-space data. $E$ represents the operator combining the undersampling mask $M$, Fourier transform $F$, and updated sensitivity maps $\hat{C}$. $S(\hat{x};\theta)$ denotes the single-coil based deep neural network with parameters $\theta$. 
We separate the optimisation into conjugate gradient SENSE~\cite{cg_SENSE} reconstruction and neural network-based reconstruction, iteratively updating $\hat{x}$.  
    \begin{equation}
\hat{x}_{rec} = (E^{H}E+\lambda I)^{-1}(E^{H}y+\lambda S(x;\theta))\label{eqn:CG_SENSE}
  \end{equation}
where $\hat{x}_{rec}$ is calculated with fixed $\theta$ parameters in the network.

An additional CSM update module is integrated to improve the original $C$ under the supervision of $\hat{x}_{rec}$ for a better SENSE reconstruction. The CSM $C$ initialised by iterative method~\cite{CSMGenerate} works as a warm start:
\begin{equation}
\hat{C} = N(C;\beta)
\end{equation}
 $N(C;\beta)$ is the network to update the CSM with parameters of $\beta$. It consisted of four single-scale convolutional layers with kernel size of 3$\times$3 followed by ReLU and a fifth layer with only 2D convolution. 

\subsection{Assessments}
The results reported in this work were computed by comparing the model outputs with fully-sampled reconstructions on a fixed test set. We reported quantitative metrics including root-mean-square-error (RMSE), normalised mean-square-error (NMSE), peak signal-to-noise ratio (PSNR), and structural similarity index (SSIM). To assess the impact on the downstream mapping task, these metrics were also calculated for the estimated T1 and T2 parameter maps.

To match the evaluation protocol used in the CMRxRecon challenge, the reconstructed images and parameter maps were cropped to a region of interest before metric computation. The quantitative assessment was specifically focused on the most clinically relevant region by retaining only the central 50\% of rows and central 33\% of columns.

\section{Results}
\subsection{Single-coil reconstruction}
To evaluate the proposed model, we conducted experiments for both the T1 and T2 mapping tasks using acceleration factors 4, 8, and 10. In this section, we compared two model configurations:
\begin{itemize}
    \item U-Net: A baseline U-Net architecture that serves as a standard encoder-decoder network.
    \item U-Net + Latent Transformer: The proposed model combining the baseline encoder-decoder model with the latent-transformer module to exploit inter-frame dependencies.
\end{itemize}

Table~\ref{tab:results-sc} summarises the results across both tasks for the three considered acceleration factors. Results demonstrate the ability of the latent transformer to effectively model the temporal correlations and improve the reconstruction.

\begin{figure}[tbh]
    \centering
    \includegraphics[width=\linewidth]{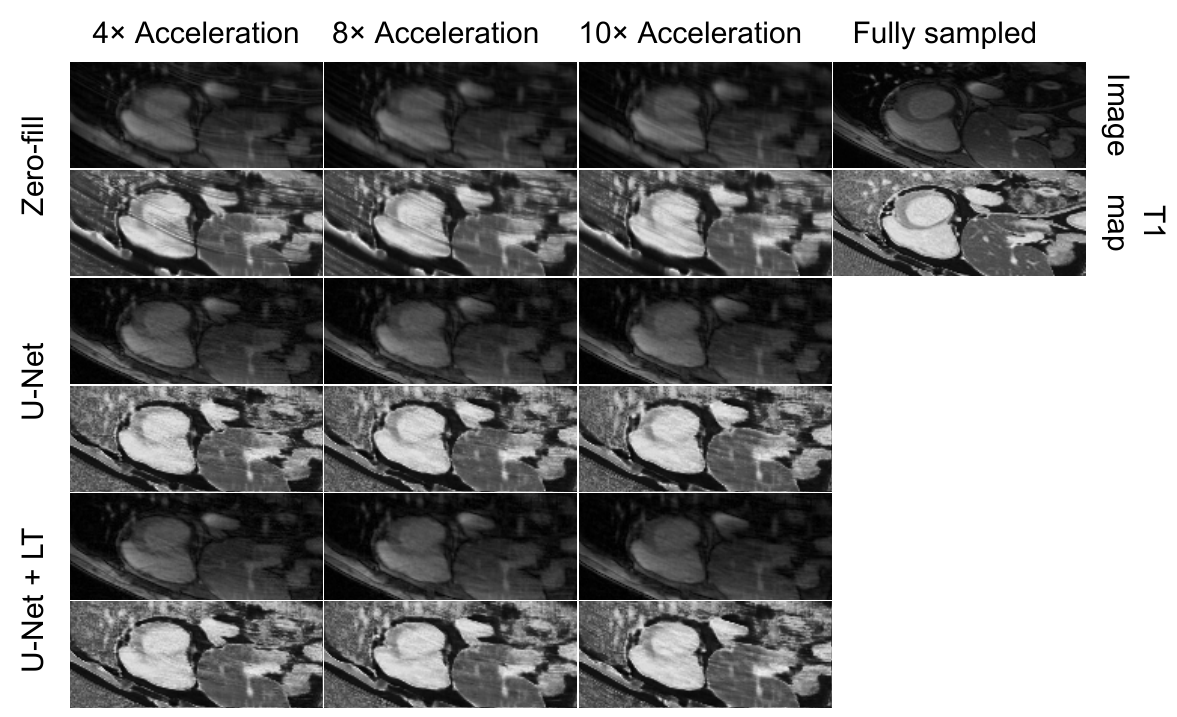}
    \caption{Qualitative comparison between zero-filling reconstruction, U-Net-based reconstruction and the proposed U-Net + LT model. The figure shows both the reconstructed images and the T1 mapping associated with the shown slice.}
    \label{fig:example-image-sc}
\end{figure}

Figure \ref{fig:example-image-sc} also qualitatively compares the reconstruction produced using Zero-filling, a U-Net model and the proposed U-Net + LT model.
\begin{table}[tbh]
\centering
\resizebox{0.9\textwidth}{!}{%
\begin{tabular}{@{}l|rrrr@{}}
\toprule
\textbf{Model} & \textbf{PSNR ↑} & \textbf{SSIM ↑} & \textbf{NMSE ↓} & \textbf{RMSE ↓} \\ \midrule
U-Net T1 4× & 30.160 & 0.811 & 0.028 & \underline{4.35E-05} \\
U-Net T1 8× & 28.158 & 0.766 & 0.044 & 5.58E-05 \\
U-Net T1 10× & 27.661 & 0.762 & 0.048 & 5.91E-05 \\
U-Net T2 4× & 28.940 & 0.827 & 0.023 & 2.81E-05 \\
U-Net T2 8× & 27.350 & 0.802 & 0.032 & 3.44E-05 \\
U-Net T2 10× & \underline{27.246} & \underline{0.808} & \underline{0.032} & \underline{3.50E-05} \\ \midrule
U-Net T1 4× + LT & \underline{30.184} & \underline{0.813} & 0.028 & 4.36E-05 \\
U-Net T1 8× + LT & \underline{28.431} & \underline{0.774} & \underline{0.040} & \underline{5.39E-05} \\
U-Net T1 10× + LT & \underline{27.933} & \underline{0.769} & \underline{0.045} & \underline{5.71E-05} \\
U-Net T2 4× + LT & \underline{29.067} & \underline{0.831} & \underline{0.022} & \underline{2.76E-05} \\
U-Net T2 8× + LT & \underline{27.469} & \underline{0.806} & \underline{0.031} & \underline{3.39E-05} \\
U-Net T2 10× + LT & 27.210 & 0.807 & 0.033 & 3.51E-05 \\ \bottomrule
\multicolumn{5}{l}{} \\ \toprule
\textbf{Model} & \textbf{Map PSNR ↑} & \textbf{Map SSIM ↑} & \textbf{Map NMSE ↓} & \textbf{Map RMSE ↓} \\ \midrule
U-Net T1 4× & 19.709 & 0.600 & 0.024 & 134.219 \\
U-Net T1 8× & 18.593 & 0.534 & 0.031 & 152.737 \\
U-Net T1 10× & 18.548 & 0.524 & 0.031 & 153.266 \\
U-Net T2 4× & 11.884 & 0.433 & 0.505 & 64.175 \\
U-Net T2 8× & 11.838 & 0.406 & 0.511 & 64.483 \\
U-Net T2 10× & 11.860 & 0.410 & 0.508 & 64.350 \\ \midrule
U-Net T1 4× + LT  & \underline{19.834} & \underline{0.607} & \underline{0.023} & \underline{132.514} \\
U-Net T1 8× + LT  & \underline{18.778} & \underline{0.538} & \underline{0.029} & \underline{149.393} \\
U-Net T1 10× + LT & \underline{18.592} & \underline{0.526} & \underline{0.030} & \underline{152.493} \\
U-Net T2 4× + LT  & \underline{11.910} & \underline{0.439} & \underline{0.502} & \underline{63.977} \\
U-Net T2 8× + LT  & \underline{11.898} & \underline{0.413} & \underline{0.504} & \underline{64.056} \\
U-Net T2 10× + LT & \underline{11.903} & \underline{0.414} & \underline{0.503} & \underline{64.017} \\ \bottomrule
\multicolumn{5}{l}{} \\
\end{tabular}%
}
\caption{Single-coil model results for both the reconstructed MR acquisitions and the computed T1 and T2 maps. The table compares a U-Net based artifact-removal process with a pipeline that used a Latent Transformer-enhanced U-Net at its core. Our report underlined the best model for a given mapping task and acceleration factor.}
\label{tab:results-sc}
\end{table}
\subsection{Multi-coil reconstruction}
To evaluate the proposed model for multi-coil MRI reconstruction, we compared two model configurations built upon MoDL~\cite{MoDL} framework:
\begin{itemize}
    \item MoDL: The baseline MoDL model using fixed coil sensitivity maps and a standard single-scale network architecture. This serves as the standard MoDL implementation.
    \item MoDL + Proposed Model: An enhanced MoDL pipeline incorporating our proposed single-coil model architecture with latent transformers and learnable coil sensitivity maps.
\end{itemize}
Experiments were conducted on the multi-coil T1 and T2 mapping datasets. Quantitative metrics compare the two MoDL-based approaches to analyse the benefits of the proposed model enhancements, including the latent transformer's ability to exploit inter-frame dependencies.

The results in Table \ref{tab:results-mc} summarise the reconstruction performance for the two models across both mapping tasks. We observed consistent improvements from the proposed techniques for integrating learnable coil sensitivity estimation and advanced single-coil modelling into the MoDL framework.
\begin{table}[tbh]
\centering
\resizebox{0.9\textwidth}{!}{%
\begin{tabular}{@{}l|rrrr@{}}
\toprule
\textbf{Model} & \textbf{PSNR ↑} & \textbf{SSIM ↑} & \textbf{NMSE ↓} & \textbf{RMSE ↓} \\ \midrule
MoDL T1 4× & \underline{34.790} & \underline{0.894} & \underline{0.026} & \underline{2.85E-05} \\
MoDL T1 8× & \underline{31.534} & \underline{0.855} & \underline{0.025} & \underline{3.80E-05} \\
MoDL T1 10× & \underline{29.904} & \underline{0.840} & \underline{0.030} & \underline{4.57E-05} \\
MoDL T2 4× & \underline{33.896} & \underline{0.911} & \underline{0.010} & \underline{1.63E-05} \\
MoDL T2 8× & \underline{29.937} & \underline{0.870} & \underline{0.018} & \underline{2.54E-05} \\
MoDL T2 10× & \underline{29.205} & \underline{0.867} & \underline{0.021} & \underline{2.79E-05} \\ \midrule
MoDL T1 4× + LT & 34.460 & 0.887 & 0.028 & 2.97E-05 \\
MoDL T1 8× + LT & 30.763 & 0.837 & 0.028 & 4.15E-05 \\
MoDL T1 10× + LT & 29.762 & 0.839 & 0.032 & 4.67E-05 \\
MoDL T2 4× + LT & 32.866 & 0.899 & 0.011 & 1.83E-05 \\
MoDL T2 8× + LT & 28.853 & 0.847 & 0.023 & 2.89E-05 \\
MoDL T2 10× + LT & 28.005 & 0.844 & 0.028 & 3.22E-05 \\ \bottomrule
\multicolumn{5}{l}{} \\ \toprule
\textbf{Model} & \textbf{Map PSNR ↑} & \textbf{Map SSIM ↑} & \textbf{Map NMSE ↓} & \textbf{Map RMSE ↓} \\ \midrule
MoDL T1 4× & \underline{22.346} & 0.719 & 0.017 & \underline{105.874} \\
MoDL T1 8× & 19.934 & 0.623 & \underline{0.021} & 131.508 \\
MoDL T1 10× & 20.004 & 0.611 & 0.022 & 129.568 \\
MoDL T2 4× & \underline{12.365} & \underline{0.577} & 0.455 & \underline{60.838} \\
MoDL T2 8× & 11.968 & 0.480 & 0.497 & 63.645 \\
MoDL T2 10× & 11.946 & 0.473 & 0.499 & 63.805 \\ \midrule
MoDL T1 4× + LT & 22.035 & \underline{0.739} & \underline{0.015} & 110.417 \\
MoDL T1 8× + LT & \underline{20.413} & \underline{0.647} & 0.023 & \underline{124.659} \\
MoDL T1 10× + LT & \underline{20.012} & \underline{0.613} & 0.022 & \underline{129.470} \\
MoDL T2 4× + LT & 12.247 & 0.547 & \underline{0.447} & 61.585 \\
MoDL T2 8× + LT & \underline{12.114} & \underline{0.494} & \underline{0.481} & \underline{62.617} \\
MoDL T2 10× + LT & \underline{11.977} & \underline{0.483} & \underline{0.496} & \underline{63.532} \\ \bottomrule
\multicolumn{5}{l}{}
\end{tabular}%
}
\caption{Multi-coil model results for both the reconstructed MR acquisitions and the computed T1 and T2 maps. The table compares a standard MoDL model with our proposed method for artifact-removal. Our report underlined the best model for a given mapping task and acceleration factor.}
\label{tab:results-mc}
\end{table}
\section{Discussion}
The introduction of the Latent Transformer (LT) module demonstrates clear improvements for the vast majority of single coil results across all analysed metrics, tasks, and acceleration factors, for both images and derived T1 and T2 maps (Table~\ref{tab:results-sc}). For multi-coil data, the improvements from the LT blocks are more modest. In particular, the LT addition degrades performance on reconstructed images but improves results for the computed T1 and T2 maps (Table~\ref{tab:results-sc}). The difference in behaviour between the two tasks likely arises from two concurring causes. First, the network works as a regularisation term. The weighting between the data consistency layer of the CG-SENSE and network may even downgrade for heavy networks. When the network becomes too complex or contains too many parameters, it may prioritise fitting the training data over maintaining consistency with the acquired data, leading to smaller weighting and reduced effect of the LT module. This can result in degraded performance and lower accuracy in image reconstruction. From Table \ref{tab:results-mc}, the proposed method outperform the original MoDL at higher acceleration factors, indicating the potential of the networks correcting for severe undersampling artifacts. Lighter variants on U-net with attention across channels may be taken into consideration. For the CSM update module, we also tried using the CSM generated on the fully-sampled k-space as a hard constrain to supervise, but got inferior performance. Networks with unrolled manner~\cite{wang2022faithful} or an additional J-SENSE module~\cite{arvinte2021deep} can be incorporated to get a better data consistency performance. Second, the LT module seems to produce some image artifacts but ultimately captures inter-frame dynamics better than simpler models focused solely on de-noising, as evidenced by improved T1 and T2 map estimates. In summary, the proposed LT framework demonstrates clear utility in exploiting temporal correlations, especially for single coil acquisitions, further validating its use for reconstructing highly accelerated MRI data for T1 and T2 mapping.


\section{Conclusion}
This work proposes a deep learning approach for reconstructing undersampled MRI that incorporates our novel Latent Transformers module to model inter-frame dependencies. Experiments on accelerated cardiac T1/T2 mapping show improved image quality and parameter mapping compared to baseline models, demonstrating the importance of temporal modelling. While promising, limitations remain including the basic U-Net architecture used. For future work, we will explore integrating the latent transformers into more advanced models like Restormer~\cite{Restormer} and unrolled networks~\cite{DCNN} with better CSM estimation to further boost performance. Overall, this study validates explicitly modelling time correlations with transformers to enable accurate reconstructions from highly accelerated quantitative MRI.

\section{Acknowledgement}
We want to show our gratitude to Zimu Huo, who provided support and advice on multi-coil reconstruction. 
This work was supported by the British Heart Foundation (RG/19/1/34160). Guang Yang was supported in part by the ERC IMI (101005122), the H2020 (952172), the MRC (MC/PC/21013), the Royal Society (IEC/NSFC/211235), the NVIDIA Academic Hardware Grant Program, the SABER project supported by Boehringer Ingelheim Ltd, and the UKRI Future Leaders Fellowship (MR/V023799/1). Fanwen Wang was supported by the UKRI Future Leaders Fellowship (MR/V023799/1). Michael T\"{a}nzer was supported by the UKRI CDT in AI for Healthcare (EP/S023283/1).

%
\clearpage
\bibliographystyle{splncs04}
\bibliography{ref}
%
\end{document}